\documentclass[article]{IEEEtran}
\usepackage{amsmath}
\usepackage{tikz}
\usepackage{cite}
\usepackage{circuitikz}
 \usepackage{tkz-euclide}
 \usetikzlibrary{angles}
 \usetikzlibrary{positioning}
\usepackage{tikz-3dplot}
\usepackage[utf8]{inputenc}
\usepackage{color}
\usepackage{lipsum}
\usepackage{soul}

\newcommand{\M}[1]{\mathbf{#1}}
\newcommand{\T}[1]{\mathrm{#1}}
\newcommand{\V}[1]{\boldsymbol{#1}}

\renewcommand{\u}[1]{\boldsymbol{\hat{#1}}}
\usetikzlibrary{shapes.symbols}

\newcommand{\Zop}{\V{\mathcal{Z}}}

\title{Sparsity of Radiating Characteristic Modes on Infinite Periodic Structures}
\author{Kurt Schab, \IEEEmembership{Member, IEEE}
\thanks{Approved for release SAND2021-7822 O, July 1 2021.}
\thanks{K. Schab is with the Department of Electrical and Computer Engineering, Santa Clara University, Santa Clara, CA, USA (e-mail: kschab@scu.edu).}
}
\begin{document}


\maketitle

\begin{abstract}
    Characteristic modes on infinite periodic structures are studied using spectral dyadic Green's functions.  This formulation demonstrates that, in contrast to the modal analysis of finite structures, the number of radiating characteristic modes is limited by unit cell size and incident wavevector (i.e., scan angle or phase shift per unit cell).  The reflection tensor is decomposed into modal contributions from radiating modes, indicating that characteristic modes are a predictably sparse basis in which to study reflection phenomena.
\end{abstract}

\begin{IEEEkeywords}
Frequency selective surfaces, periodic systems, Floquet expansion, electromagnetic scattering
\end{IEEEkeywords}

\section{Introduction}

\IEEEPARstart{C}{haracteristic} mode analyses \cite{Garbacz_TCMdissertation,Harrington_1971a,Harrington_1971b} of arrays and metasurfaces typically follow one of two trends aimed at reducing what is, in principle, a problem of infinite spatial extent to one of finite size.  The simplest of these approaches is to analyze large, finite periodic structures as a single object e.g., \cite{li2018metasurface,zhao2018characteristic,lin2018truncated,lin2019recent,genovesi2021characteristic}.  This method involves no changes to the underlying characteristic mode formulation but yields current distributions which depend significantly on the size and shape of the finite structure being studied, e.g., modes of a large rectangular array resemble the modes of a large rectangular plate \cite{KingAJ_PhDThesis}.  Though local whole domain basis functions defined over each unit cell may be used to accelerate computation \cite{Tzanidis_2012a, KingAJ_PhDThesis, Lonskyetal_CMofDipoleArrays,cheng2020development}, these analyses are typically limited by the computational cost of modeling electrically large systems, making asymptotic analysis toward infinite systems prohibitively complex.  In contrast, another approach involves studying the modal behavior of a single unit cell, either in isolation \cite{chukwuka2020stored} or through the use of periodic boundary conditions \cite{maalik2016characteristic,haykir2018characteristic,Haykir_2019a,Angiulli_2000a,guo2020miniaturized,wu2021broadband} or their approximation \cite{ethier2012antenna}.  The use of periodic boundary conditions appears to be promising as it naturally captures inter-element mutual coupling effects while requiring a relatively low number of degrees of freedom in representing fields and currents, however previous work in this area is largely limited to the numerical analysis of surfaces under normal illumination, a case where familiar properties of the method of moments impedance matrix (e.g., transpose symmetry) hold and no significant changes to the characteristic mode formulations are required.

In this letter, the characteristic modes of infinite periodic structures are formulated using spectral Green's functions, i.e., Floquet mode expansion (numerically studied in \cite{angiulli2000application} and qualitatively discussed in \cite{Haykir_2019a}), in order to theoretically examine properties not present in the analysis of finite structures.  We focus on the study of periodic structures constructed from patterned perfectly conducting (PEC) screens in free space.  This restriction leads to several interesting analytic results related to the fact that only a small number of radiating characteristic modes exist, a property made clear by analysis of the spectral Green's function for these particular problems.  Similar results hold for screens supported by lossless dielectric layers.  Additionally, modal decomposition of a surface's reflection tensor is discussed, which, due to the limited number of radiating modes, is found to be an extremely sparse representation of this figure of merit.

\section{Spectral representations}
\label{sec:spectral}

\begin{figure}
    \centering
	\tdplotsetmaincoords{60}{120}
    \begin{tikzpicture}[tdplot_main_coords,scale=0.9,transform shape]
			\draw[fill = yellow!10] (0,0,0) -- (0,1,0) -- (1,1,0) -- (1,0,0) -- (0,0,0);
			\begin{scope}[shift={(2,0)}]
			\draw[fill = yellow!10] (0,0,0) -- (0,1,0) -- (1,1,0) -- (1,0,0) -- (0,0,0);
			\end{scope}
			\node at (2.5,0.5) {$\varOmega$};
			\begin{scope}[shift={(2,2)}]
			\draw[fill = yellow!10] (0,0,0) -- (0,1,0) -- (1,1,0) -- (1,0,0) -- (0,0,0);
			\end{scope}
			\begin{scope}[shift={(0,2)}]
			\draw[fill = yellow!10] (0,0,0) -- (0,1,0) -- (1,1,0) -- (1,0,0) -- (0,0,0);
			\end{scope}	
			\draw[dashed] (-1,-0.5,0) -- (4,-0.5,0);
			\draw[dashed] (-1,1.5,0) -- (4,1.5,0);
			\draw[dashed] (-1,3.5,0) -- (4,3.5,0);
			\draw[dashed] (-0.5,-1,0) -- (-0.5,4,0);
			\draw[dashed] (1.5,-1,0) -- (1.5,4,0);
			\draw[dashed] (3.5,-1,0) -- (3.5,4,0);
			\draw[->] (0.5,2.5) -- (0.5,3.85) node[right] {$x$};
			\draw[->] (0.5,2.5) -- (-0.95,2.5) node[right] {$y$};
			\draw[<->] (1.5,-1) -- node[above left] {$T_y$} (3.5,-1);
			\draw[<->] (4,-0.5) -- node[below left] {$T_x$} (4,1.5);
    \end{tikzpicture}
    \caption{Four unit cells within an infinite periodic structure lying in the $xy$ plane.  Each cell is of dimension $T_x\times T_y$ and contains an identical conducting patch $\varOmega$.  These patches may be connected across unit cell boundaries.}
    \label{fig:schem}
\end{figure}
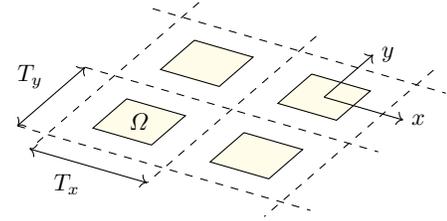

The characteristic mode analysis of periodic infinite structures is made relatively simple through the use of a spectral decomposition of currents existing within each unit cell.  This process is reviewed over the next two sections in order to highlight a few of its key features relevant to characteristic mode analysis.  Readers should consult more comprehensive references on this topic, e.g., \cite{pozar1984analysis,pozar1987radiation,cwik1987scattering,mittra1988techniques,jin2011theory,munk2005frequency}, for further details.

Consider a planar, rectangular unit cell $S$ of dimension $T_x \times T_y$ containing a perfectly conducting patch $\varOmega$ lying in the $z=0$ plane, depicted in Fig.~\ref{fig:schem}.  Assume this unit cell repeats infinitely in both $x$ and $y$ directions.  A global current density $\V{J}(\V{r})$ in this plane may be expanded as
\begin{equation}
    \V{J}(\V{r}) = \V{j}(\V{r})\T{e}^{\T{j}\V{k}_\T{i}\cdot\V{r}}
\end{equation}
where $\V{j}(\V{r})$ is the local current density repeated identically in each unit cell and $\V{k}_\T{i}$ represents the tangential component of an incident wavevector (i.e., scan angle or prescribed phase shift per unit cell).  Due to its periodicity, the current density $\V{j}$ may be expanded into a two dimensional Fourier basis (Floquet modes) using the transform pair
\begin{equation}
    \V{j}(\V{r}) = \sum_\gamma \tilde{\V{\jmath}}_\gamma \T{e}^{\T{j}\V{k}_{\gamma}\cdot\V{r}}
    \label{eq:j-periodic-floquet}
\end{equation}
and
\begin{equation}
    \V{\tilde{\jmath}}_{\gamma} = \frac{1}{T_xT_y}\int_S \V{j}(\V{r}) \T{e}^{-\T{j}\V{k}_{\gamma}\cdot\V{r}}\T{d}\V{r}
\end{equation}
where
\begin{equation}
    \V{k}_{\gamma} = \u{x}\frac{2\pi p}{T_x} + \u{y}\frac{2\pi q}{T_y}
\end{equation}
and the indices $p$ and $q$ are contained within the meta-index~$\gamma$.  Each spectral current component $\tilde{\V{\jmath}}_\gamma$ produces a phase matched tangential electric field $\tilde{\V{e}}_\gamma$ via a spectral Green's function.  Hence a spectral impedance operator\footnote{Following \cite{Harrington_1971a}, the impedance operator maps a current onto the negated tangential scattered field it produces.  Hence, $\tilde{\V{e}}_\gamma$ should be interpreted as only the tangential field component.} acting on each spectral current component is given by
\begin{equation}
    \tilde{\V{e}}_\gamma = -\Zop_\gamma \tilde{\V{\jmath}}_\gamma
\end{equation}
where
\begin{equation}
    \Zop_\gamma = \frac{\eta}{2kk'_{\gamma z}}
	\begin{bmatrix}
	k^2-k'^2_{\gamma x} & - k'_{\gamma x}k'_{\gamma y} \\
	- k'_{\gamma x}k'_{\gamma y} & k^2-k'^2_{\gamma y} \\
	\end{bmatrix}
	= (k'_{\gamma z})^{-1}\V{\mathcal{A}}_{\gamma}
\end{equation}
with $\V{\mathcal{A}}_{\gamma}$ being a real-valued dyadic, the free space wavenumber $k = \omega\sqrt{\mu_0\varepsilon_0}$, and the primed wavevector given by
\begin{equation}
	\V{k}'_\gamma = \V{k}_\T{i} + \V{k}_{\gamma} + \u{z}\sqrt{k^2 - |\V{k}_\T{i} + \V{k}_{\gamma}|^2}.
	\label{eq:kpqz}
\end{equation}

\begin{figure}
    \centering
    \includegraphics[width=3.1in]{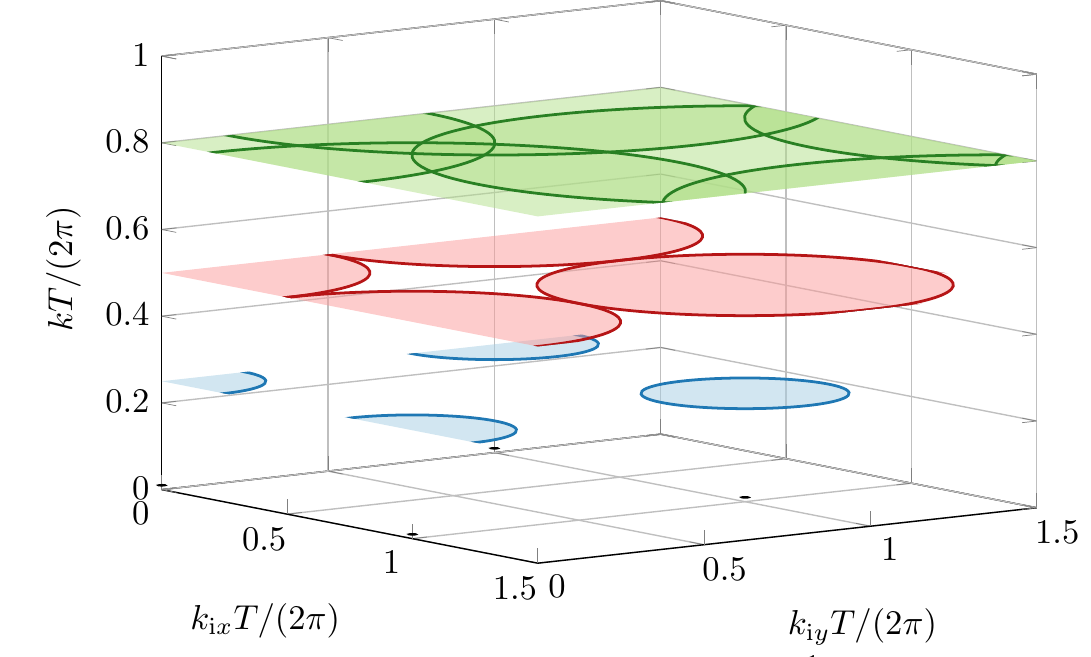}
    \caption{Zones indicating the presence of radiating spectral components for a square unit cell of dimension $T$.  Three cuts through $\V{k}_\T{it}$ space are shown at $kT/(2\pi) = T/\lambda = 0.25$ (blue), $0.5$ (red), and $0.8$ (green).}  
    \label{fig:zones}
\end{figure}

By the expansion in \eqref{eq:j-periodic-floquet}, the total periodic tangential field in the $z=0$ plane is made up of many spectral contributions
\begin{equation}
    \V{e}(\V{r}) = \sum_\gamma \tilde{\V{e}}_\gamma \T{e}^{\T{j}\V{k}_{\gamma}\cdot\V{r}} = -\sum_\gamma \Zop_\gamma \tilde{\V{\jmath}}_\gamma \T{e}^{\T{j}\V{k}_{\gamma}\cdot\V{r}}.
    \label{eq:ezj}
\end{equation}
Invoking Poynting's theorem over a region enclosing one unit cell, orthogonality of complex exponentials may be employed to reduce the complex power $P$ provided per unit cell to a single sum of spectral contributions $P_\gamma$, i.e.,
\begin{equation}
	P = \sum_{\gamma} \frac{T_xT_y}{2k'_{\gamma z}}\tilde{\V{\jmath}}_{\gamma}^*\cdot\V{\mathcal{A}}_{\gamma}\tilde{\V{\jmath}}_{\gamma} = \sum_{\gamma} P_{\gamma}.
\end{equation}
For any modal vector coefficient $\tilde{\V{\jmath}}_{\gamma}$ and real transverse wavenumber $\V{k}_{\gamma}$, the term $\tilde{\V{\jmath}}_{\gamma}^*\cdot\V{\mathcal{A}}_{\gamma}\tilde{\V{\jmath}}_{\gamma}$ appearing in the above expression is purely real, though indefinite. This implies that the spectral powers $P_\gamma$ are either purely radiating (real) or purely reactive (imaginary), depending on the nature of the longitudinal wavenumber $k'_{\gamma z}$, which in turn is dictated by the relative magnitudes of the transverse and total wavenumbers via \eqref{eq:kpqz}. Within a two dimensional space of incident wavenumber components $k_{\T{i}x}$ and $k_{\T{i}y}$, the radiating (i.e., visible) region for each spectral component is a circle of radius $k$ centered at
\begin{equation}
	k_{\T{i}x}^\gamma = -\frac{2\pi p}{T_x},\quad k_{\T{i}y}^\gamma = -\frac{2\pi q}{T_y} .
\end{equation}
These regions are plotted for several values of the wavenumber $k$ in Fig.~\ref{fig:zones} for a square unit cell of dimension $T$.  

\section{Spatial impedance matrices}
\label{sec:mom}

Expanding the periodic current density $\V{j}(\V{r})$ into a real-valued $N$-dimensional basis $\{\V{\psi}_\alpha(\V{r})\}$ (e.g., rooftop basis functions~\cite{cwik1987scattering}) with coefficients $\{I_\alpha\}$ and applying Galerkin testing to \eqref{eq:ezj} leads to
\begin{equation}
    V_\beta = -\frac{1}{T_xT_y}  \sum_{\alpha=1}^N I_\alpha \sum_\gamma \V{f}_{\gamma\beta}^*\cdot\Zop_\gamma \V{f}_{\gamma\alpha}
\end{equation}
where
\begin{equation}
    \V{f}_{\gamma\alpha} = \int_S \V{\psi}_\alpha(\V{r}) \T{e}^{-\T{j}\V{k}_{\gamma}\cdot\V{r}}\T{d}\V{r}
\end{equation}
and
\begin{equation}
    V_\beta = \int_S \V{e}(\V{r})\cdot\V{\psi}_\beta(\V{r})\T{d}\V{r}.
\end{equation}
Collecting $N$ such equations allows for the construction of a method of moments system
\begin{equation}
    \M{V} = -\M{Z}\M{I}.
\end{equation}
Hence we interpret the complex impedance matrix entry $Z_{\alpha\beta}$ as the sum of spectral terms $Z_{\alpha\beta,\gamma}$, each of which, in general, is complex.  Based on the discussion in Sec.~\ref{sec:spectral}, the dyadic $\V{\mathcal{Z}}_{\gamma}$ takes on purely real or purely imaginary values when the Floquet harmonic with index $\gamma$ is radiating or evanescent, i.e.,
\begin{equation}
Z_{\alpha\beta,\gamma} = R_{\alpha\beta,\gamma} = a\V{f}_{\alpha, \gamma}^*\cdot \V{\mathcal{A}}_{\gamma}\V{f}_{\beta, \gamma},\quad|\V{k}_{\gamma}+\V{k}_\T{i}|<k
\end{equation}
and
	\begin{equation}
Z_{\alpha\beta,\gamma} = \T{j}X_{\alpha\beta,\gamma} = \T{j}b\V{f}_{\alpha, \gamma}^*\cdot \V{\mathcal{A}}_{\gamma}\V{f}_{\beta, \gamma},\quad|\V{k}_{\gamma}+\V{k}_\T{i}|\geq k
\end{equation}
with real-valued scaling factors $a$ and $b$. Note that the radiating spectral contributions are Hermitian while the evanescent spectral contributions are anti-Hermitian, i.e.,
\begin{equation}
R_{\alpha\beta,\gamma} = R_{\beta\alpha,\gamma}^* ~~\text{and}~~ \T{j}X_{\alpha\beta,\gamma} = -\left(\T{j}X_{\beta\alpha,\gamma}\right)^*.
\label{eq:xabpq-hermitian}
\end{equation}
We may divide the spectrum of Floquet harmonics into two subspectra and rewrite the impedance matrix $\M{Z}$ as
\begin{equation}
\M{Z} = \sum_\T{rad}\M{R}_{\gamma} + \T{j}\sum_\T{ev}\M{X}_{\gamma} = \M{R}+\T{j}\M{X},
\end{equation}
where $\sum_\T{rad}$ and $\sum_\T{ev}$ denote sums over spectral meta-indices $\gamma$ corresponding to radiating and evanescent modes, respectively.  The matrices $\M{R}$ and $\M{X}$ are, in general, complex, however due to the conditions observed in \eqref{eq:xabpq-hermitian}, they are also Hermitian symmetric.  Therefore, the Hermitian parts $\M{R}$ and $\M{X}$ of the impedance matrix $\M{Z}$ are naturally constructed from contributions arising from the radiating and evanescent Floquet subspectra, respectively.  
 These operators maintain their direct connection to per-unit-cell radiated and reactive power, i.e.,
\begin{equation}
P_\T{r} = \frac{1}{2}\M{I}^\T{H}\M{R}\M{I}
\end{equation} 
and
\begin{equation}
P_\T{x} = 2\omega(\bar{W}^\T{m} - \bar{W}^\T{e}) = \frac{1}{2}\M{I}^\T{H}\M{X}\M{I}.
\end{equation}

The number of radiating spectral components is predetermined by the parameters $\V{k}_\T{i}$, $T_x$ and $T_y$.  This, in turn, determines the rank of the radiation and reactance matrices $\M{R}$ and $\M{X}$.  Note that resonant evanescent spectral components may alter the rank of the matrix $\M{X}$ at discrete combinations of parameters.  If the periodicity is sufficiently small that $T_{x/y}<\lambda/2$, then the visible regions of each spectral component are non-overlapping and at most only one radiating component exists at each incident spatial frequency $\V{k}_\T{i}$, cf Fig.~\ref{fig:zones}.

Modifying the problem to include a dielectric support layer with or without a conductive backing involves a change of the underlying periodic impedance operator, see \cite{pozar1984analysis,pozar1987radiation, angiulli2000application, munk2005frequency,kwon2014energy}.  In this case, radiating spectral components contribute to both the radiative and reactive parts of the impedance matrix, though the number of radiating components (and thus the rank of the matrix $\M{R}$) is still limited and determined via conditions similar to those illustrated in Fig.~\ref{fig:zones}, see \cite[Ch. 5]{munk2005frequency} and \cite{kwon2014energy}.

\section{Characteristic modes}
For perfectly conducting structures, characteristic modes diagonalize both the impedance and radiation operators \cite{Harrington_1971a}.  Using the method of moments formulation in Sec.~\ref{sec:mom}, the generalized eigenvalue problem producing characteristic modes reads \cite{Harrington_1971b}
\begin{equation}
    \M{Z}\M{I}_n = (1+\T{j}\lambda_n) \M{R}\M{I}_n,
    \label{eq:cmgep}
\end{equation}
where $\{\M{I}_n\}$ are eigenvectors corresponding to the periodic currents $\{\V{j}_n\}$ represented in a spatial basis, and $\{\lambda_n\}$ are the associated eigenvalues.  By the Hermitian symmetry of both $\M{R}$ and $\M{X}$, eigenvectors associated with non-degenerate eigenvalues are orthogonal in the sense that
\begin{equation}
	\M{I}_m^\T{H}\M{R}\M{I}_n = 0,~m\neq n,\quad \M{I}_m^\T{H}\M{X}\M{I}_n = 0,~m\neq n.
\end{equation}
Due to the limited number of radiating spectral components, there exist only $N_\T{r}$ radiating modes, where $N_\T{r} = \T{rank}~\M{R}$.  For these modes, radiated and reactive powers are related via
\begin{equation}
\M{I}_m^\T{H}\M{R}\M{I}_m = 2P_{\T{r},m},
\end{equation} 
and 
\begin{equation}
\M{I}_m^\T{H}\M{X}\M{I}_m = 2\lambda_mP_{\T{r},m}.
\label{eq:xorth}
\end{equation}
While each radiating mode may have projections onto multiple radiating spectral components (i.e., planewave radiation in multiple directions), the orthogonality properties above indicate that these modes have orthogonal far fields when integrated over all radiation directions.  Non-radiating modes (in the null space of $\M{R}$) have eigenvalues $|\lambda_n| = \infty$, save for those which are also resonant (in the null space of $\M{X}$) which have indeterminate eigenvalues.  

\begin{figure}
    \centering
    \includegraphics[width=3.25in]{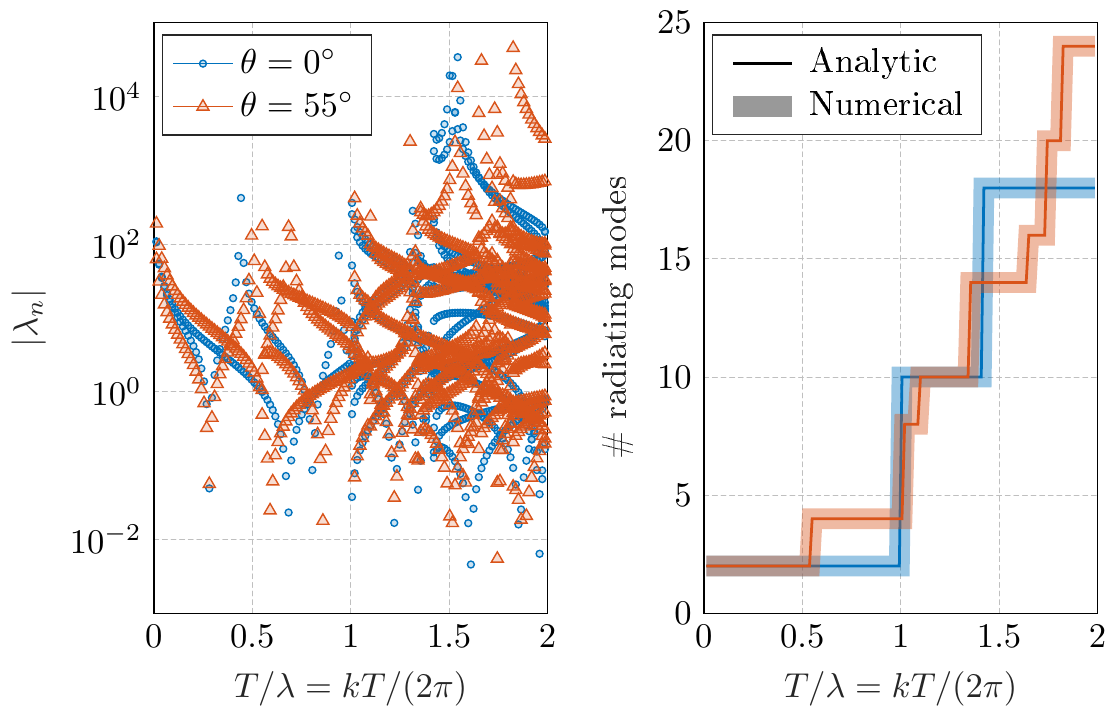}
    \caption{Characteristic mode eigenvalues (left) of a split ring resonator over a range of unit cell electrical sizes and the number of radiating modes (right) obtained by characteristic mode decomposition and analytic prediction.}
    \label{fig:modes}
\end{figure}

\begin{figure}
    \centering
    \includegraphics[width=3.5in]{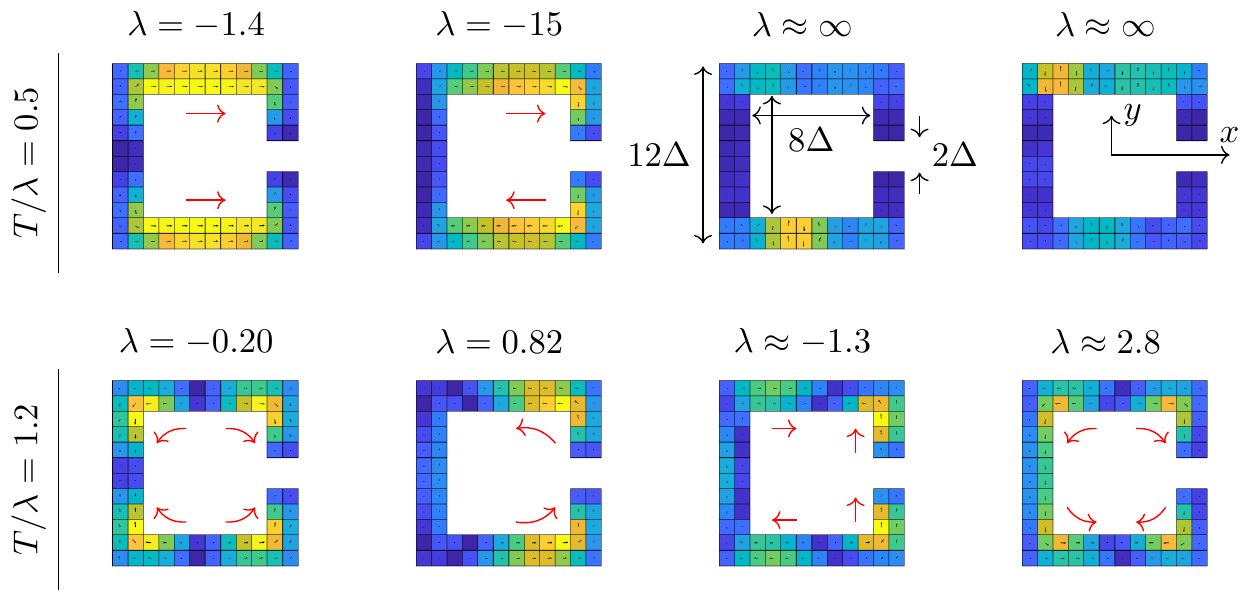}
    \caption{Numerically calculated characteristic modes for a square split ring resonator, see inset for dimensions, within a square unit cell of dimension $T = 20\Delta$.  Normal incidence is assumed ($\V{k}_\T{i} = 0$) and modal current distributions are plotted for two electrical sizes. Yellow (blue) colors indicate current maxima (minima), while red arrows clarify current direction.}
    \label{fig:currents}
\end{figure}

Numerically calculated eigenvalues for a square split ring resonator within a square unit cell of dimension $T$ are shown in Fig.~\ref{fig:modes}.  Also shown are the number of radiating modes obtained numerically and through the analytic prediction based on overlapping spectral zones, cf Fig.~\ref{fig:zones}.  Data are plotted for two incident wavevectors of the form of $\V{k}_\T{i} = \u{x}k\sin\theta$ with $\theta = 0^\circ$ and $55^\circ$, illustrating the effect of oblique incidence in lowering the electrical size required to observe higher-order radiating spectral components (i.e., grating lobes).    Modal current distributions of the modes with smallest eigenvalue magnitude are plotted in Fig.~\ref{fig:currents} for normal incidence at two frequencies below and above the introduction of grating lobes.  Non-radiating modes exhibit seemingly random current distributions with eigenvalues tending toward infinity, though they are in practice saturated at a dynamic range on the order of $10^{16}$ due to finite numerical precision. 

We now consider an incident field constructed from propagating plane waves
\begin{equation}
\V{e}_\T{i} = \sum_{\T{rad}}\tilde{\V{e}}_{\T{i}\gamma}\T{e}^{\T{j}\V{k}_{\gamma}\cdot\V{r}}\quad\Rightarrow\quad \M{V}_\T{i} = \sum_{\T{rad}} \M{f}^\T{H}_\gamma\tilde{\V{e}}_{\T{i}\gamma},
\label{eq:ei}
\end{equation}
where $\M{f}_\gamma$ is a collection of vectors $\V{f}_{\gamma\alpha}$ for all members of the chosen basis. Enforcing the PEC boundary condition canceling scattered and incident tangential fields over each patch $\varOmega$, the induced current distribution $\M{I}$ can, by \eqref{eq:cmgep}-\eqref{eq:xorth}, be expressed in terms of $N_\T{r}$ radiating characteristic modes via
\begin{equation}
\M{I} = \sum_{n=1}^{N_\T{r}} \alpha_n\M{I}_n, \quad \alpha_n =  \frac{\M{I}_n^\T{H}\M{V}_\T{i}}{2P_{\T{r},n}(1+\T{j}\lambda_n)}.
\label{eq:alpha}
\end{equation}
Note that, by \eqref{eq:ei},
\begin{equation}
    \M{I}_n^\T{H}\M{V}_\T{i} = \sum_{\T{rad}} (\M{f}_\gamma\M{I}_n)^\T{H}\tilde{\V{e}}_{\T{i}\gamma},
\end{equation}
and only characteristic modes with non-zero projection onto radiating spectral components may be excited, consistent with expectations of reciprocity.

\section{Modal reflection coefficients}

\begin{figure}
    \centering
    \includegraphics[width=3.25in]{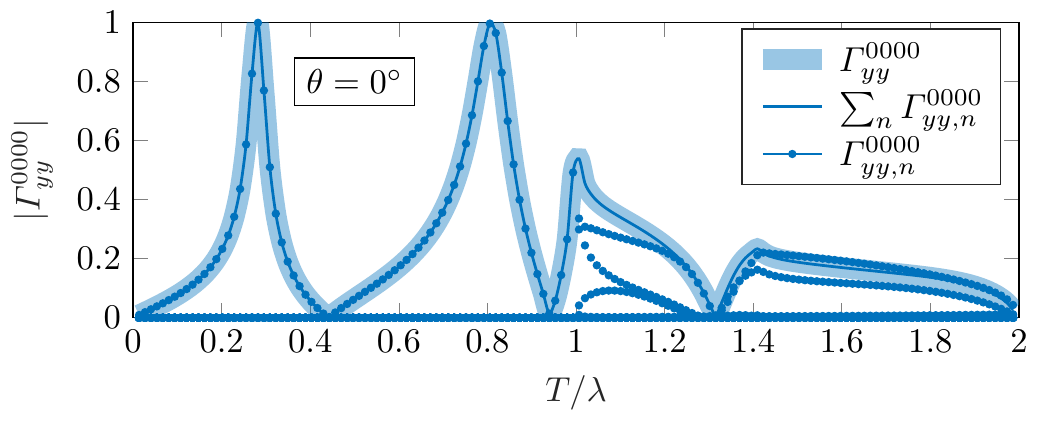}\\
    \includegraphics[width=3.25in]{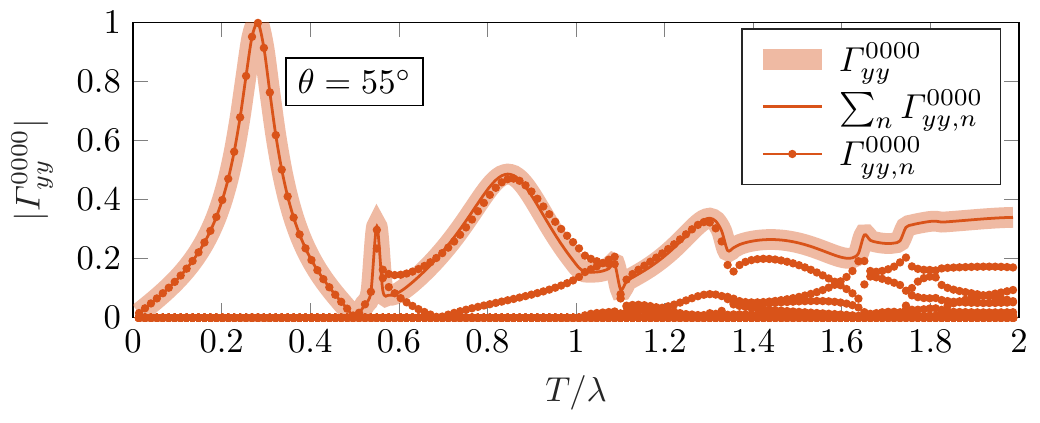}\\
    \caption{Specular co-polarized reflection coefficients $\varGamma_{yy}^{0000}$ calculated for the split ring resonator studied in Fig.~\ref{fig:currents} using an incident wavevector of the form $\V{k}_\T{i} = \u{x}k\sin\theta$ with $\theta = 0^\circ$ (top) and $\theta = 55^\circ$ (bottom).}
    \label{fig:reflection}
\end{figure}

Given an incident field described by \eqref{eq:ei}, a reflection tensor mapping the incident $\gamma'$ spectral field component into the scattered $\gamma$ spectral field component may be written as 
\begin{equation}
     \tilde{\V{e}}_{\gamma} = \sum_{\gamma'}\V{\Gamma}^{\gamma\gamma'}\tilde{\V{e}}_{\T{i}\gamma'}.
\end{equation}
By the expansions in \eqref{eq:ei} and \eqref{eq:alpha}, the reflection tensor $\V{\Gamma}^{\gamma\gamma'}$ may be decomposed into modal contributions
\begin{equation}
    \V{\Gamma}^{\gamma\gamma'} = \sum_{n=1}^{N_\T{r}} \V{\Gamma}^{\gamma\gamma'}_n = -\sum_{n=1}^{N_\T{r}} \frac{\Zop_\gamma\M{f}_\gamma\M{I}_n}{T_xT_y} \frac{\M{I}_n^\T{H}\M{f}_{\gamma'}^\T{H}}{2P_{\T{r},n}(1+\T{j}\lambda_n)},
    \label{eq:gamma-exp}
\end{equation}
where the right hand side is understood to be constructed from outer products of two-dimensional vectors.  The action of each dyad $\V{\Gamma}^{\gamma\gamma'}_n$, has the interpretation of taking the projection of the incident field component $\tilde{\V{e}}_{\T{i}\gamma'}$ onto the characteristic mode $\M{I}_n$ and applying the result as a weighting for the projection of that same characteristic mode onto the spectral component $\tilde{\V{e}}_\gamma$ of the scattered field. Longitudinal scattered fields may be reconstructed by considering the propagation vector associated with each spectral component.

In Fig.~\ref{fig:reflection}, the specular co-polarized reflection coefficient $\varGamma_{yy}^{0000}$ is plotted along with its modal contributions.  As predicted, at low frequencies at most two characteristic modes (here one, by symmetry) contribute to specular reflection.  As frequency increases and more radiating spectral components appear, more modes contribute.  Consistent with the results in Figs.~\ref{fig:zones} and \ref{fig:modes}, the appearance of radiating characteristic modes is modulated by the incident wavevector.  In all cases studied here, the modal expansions in \eqref{eq:alpha} and \eqref{eq:gamma-exp} using $N_\T{r}$ characteristic modes accurately reconstruct the driven  current and reflection coefficient to numerical precision.

\section{Discussion}

In contrast to the analysis of structures of finite size, only a finite number of radiating characteristic modes are present on the infinite periodic systems studied in this letter.  This distinction is primarily due to the nature of the spectral Green's functions used, which give rise to radiation operators of finite and predictable rank based on the number of radiating spectral components.  The number of radiating spectral components governs the number of radiating characteristic modes and is predictable \emph{a priori} by standard Floquet mode theory.  All non-radiating characteristic modes have infinite or indeterminate eigenvalues and cannot be driven by planewave excitation.  This result indicates that, assuming excitations of the form of propagating plane waves, expansion of the driven current in characteristic modes is rigorously sparse, as is the expansion of the reflection tensor $\V{\Gamma}^{\gamma\gamma'}$.  Inclusion of lossless dielectric support layers leads to similar results.

Additionally, it is clear from the analysis of the cases considered here that the Hermitian parts, not the element-wise real and imaginary parts, of the impedance matrix are of interest in constructing the characteristic mode eigenvalue problem.  This distinction is of particular importance in cases involving oblique incidence (non-zero phase shift per unit cell) where the impedance matrix is not transpose symmetric.

With the unique features of characteristic modes on infinite and periodic systems come several interesting challenges and opportunities for future research.  Similar to problems involving driven antenna systems \cite{yee1973self}, it is unclear if or how characteristic modes could accurately model currents excited by localized feeds with incident fields containing evanescent spectral components.  Modal tracking (see \cite{Masek_2020a} for background and references) will also involve new challenges, with the single frequency parameterization now elevated to a three-dimensional space of frequency and incident wavevector components.  There the interpretation of symmetry, common in frequency selective surfaces and reflectarrays \cite{munk2005frequency}, and its impact on tracking must be reassessed.  Additionally, the strictly sparse nature of characteristic modes in this context may be a significant aid in formulating and solving optimization problems used to calculate physical bounds on certain classes of electromagnetic systems, e.g.,~\cite{ludvig2019fundamental,ludvig2020physical,2020_Gustafsson_NJP}.

\section*{Acknowledgements}

This work was funded by Sandia National Laboratories, a multimission laboratory managed and operated by NTESS, LLC, a wholly owned subsidiary of Honeywell International, Inc., for the U.S. Department of Energy’s NNSA under contract DE-NA0003525. This paper describes objective technical results and analysis. Any subjective views or opinions that might be expressed in the paper do not necessarily represent the views of the U.S. DoE or the U.S. Government.

The author thanks John Borchardt at Sandia National Laboratories for insightful discussions related to this work.

\bibliographystyle{IEEEtran}
\bibliography{main}

\end{document}